\documentclass[12pt]{article}
\usepackage{graphicx}
\begin{document}
\rightline{NKU-01-SF2}
\bigskip

\begin{center}
{\Large\bf Charged  Black Hole Solutions
 in Einstein-Born-Infeld gravity with a Cosmological Constant

}

\end{center}
\hspace{0.4cm}
\begin{center}
{Sharmanthie Fernando* \footnote{fernando@nku.edu} and Don Krug** \footnote{krugd@nku.edu}}\\
{\small\it *Department of Physics \& Geology}\\
{\small\it **Department of Mathematics \& Computer Science}\\
{\small\it Northern Kentucky University}\\
{\small\it Highland Heights}\\
{\small\it Kentucky 41099}\\
{\small\it U.S.A.}\\

\end{center}

\begin{center}
{\bf Abstract}
\end{center}

\hspace{0.7cm}{\small 

We construct black hole solutions to 
Einstein-Born-Infeld 
gravity with a cosmological constant.
Since an elliptic function appears in the solutions for the metric, we construct horizons numerically. The causal structure of these solutions differ drastically from their counterparts in Einstein-Maxwell gravity with a cosmological constant. The charged de-Sitter black holes can have  up to three horizons and the charged anti-de Sitter black hole can have one or two depending on the parameters chosen.
}
\newline
\newline
{\it Key words}: Born-Infeld, Cosmological constant, Black Holes

\section{Introduction}

Born-Infeld electrodynamics was first introduced in 1930's to obtain a finite energy density model for the electron \cite{born}. It has attracted considerable interest in recent times due to various reasons. One of the motivations being the observations that it arises naturally in open  superstrings and in D-branes \cite{leigh}. The low energy effective action for an open superstring in loop calculations lead to Born-Infeld type actions \cite{frad}. It has also being observed that the Born-Infeld action arises as an effective action governing the dynamics of vector-fields on D-branes \cite{tsey}. For a  review of aspects of Born-Infeld theory in string theory see Gibbons \cite{gib}.

In recent times, the cosmological constant has been considered for several theoretical and observational reasons. One of the strongest supports have  been the recent results from a supernova which suggests a positive cosmological constant \cite{carroll}. From the theoretical point of view, recent advances in string theory have been subjected to studies in all dimensions with a negative cosmological constant \cite{malde}. Asymptotically anti-de Sitter black holes are more interesting than their counterparts since they allow more geometries for black hole horizons than for the $\Lambda =0$ case \cite{van}. Also they are thermodynamically stable which has been an inspiration for much  recent work \cite{page}. Therefore studying black holes in (anti)de Sitter spaces has an important place in current research.

In this paper we explore the existence of electrically charged black holes in Einstein-Born-Infeld gravity with a cosmological constant for a U(1) gauge field. Our motivation is to see how the singular nature of  black holes gets modified when coupled to  non-linear electrodynamics. We find that  depending on the value of the cosmological constant $\Lambda$, mass parameter $M$, coupling constant $\beta$ and the charge $Q$, the black hole solutions can have up to three horizons.

The paper is organized as follows: In section 2 the basic equations are derived. In section 3 the general solutions to static spherically symmetric solutions are constructed. In section 4 the solutions are studied in detail for various values of the parameters of the theory. In section 5 the temperature is calculated and finally in section 6 the conclusion are given.

\section{Basic Equations}
In this section we will derive the equations of motion for non-linear electrodynamics. The most general action for such a theory coupled to gravity with a cosmological constant is as follows:
\begin{equation}
S = \int d^4x \sqrt{-g} \left[ \frac{(R - 2 \Lambda)}{16 \pi G} + L(F) \right]
\end{equation}
Here, $L(F)$ is a function of the field strength $F_{\mu \nu}$ only. For the  weak field limit, $L(F)$ has to be of the form 
\begin{equation}
L(F) = - F^{\mu \nu}F_{\mu \nu} + O(F^4)
\end{equation}
In this paper, we will study a particular non-linear electrodynamics called Born-Infeld theory which has attracted lot of attention due to its relation to string effective actions. The function $L(F)$ for Born-Infeld may be expanded to 
\begin{equation}
L(F) = 4 \beta^2 \left( 1 - \sqrt{ 1 + \frac{ F^{\mu \nu}F_{\mu \nu}}{ 2 \beta^2}} \right)
\end{equation}
Here, $\beta$ has dimensions $length^{-2}$ and $G$  $length^2$. In the following sections we will take $16 \pi G = 1$.

By extremising the Lagrangian in eq.(1), with respect to the metric $g_{\mu \nu}$
and the electrodynamics potential $A_{\mu}$, one obtains the corresponding field equations as follows:
\begin{equation}
\bigtriangledown_{\mu}\left(\frac{F^{\mu \nu}}{ \sqrt{ 1 + F^2 / 2 \beta^2}} \right)=0
\end{equation}
\begin{equation}
R_{\mu \nu} - g_{\mu \nu} \Lambda = \left( \frac{\partial L(F)}{\partial g^{\mu \nu} } + \frac{g_{\mu \nu} A(F)}{2} \right)
\end{equation}
where 
\begin{equation}
A(F) = -L(F) + \frac{\partial L(F)}{\partial g^{ab}} g^{ab}; \hspace{1.0cm}
\frac{\partial L(F)}{\partial g^{\mu \nu}} =  \frac{-2 F_{\beta \nu}F^{\beta}_{\mu}}{\sqrt{1 + F^2/2\beta^2}}
\end{equation}

\section{Static Spherically Symmetric Solutions}
We will study  static spherically symmetric charged solutions to Einstein-Born-Infeld gravity with a cosmological constant. The most general metric for such a  configuration can be written as
\begin{equation}
ds^2 = - e^{2\mu} dt^2 + e^{2\nu} dr^2 + r^2 ( d \theta^2 + sin^2 \theta d \phi^2)
\end{equation}
To simplify the equations, we use tetrad formalism and Cartan structure equations.
The obvious orthonomal basis  is the following:
\begin{equation}
\theta^0 = e^{\mu} dt; \hspace{1.0cm} \hspace{1.0cm} \theta^1 = e^{\nu} dr; \hspace{1.0cm} \theta^2 =  r d\theta; \hspace{1.0cm} \theta^3 = r sin(\theta) d \phi
\end{equation}
The indices $a,b = 0,1,2,3$ are for the orthonomal basis and $\mu, \nu = 0,1,2,3,$ for the coordinate basis with $x^0 =t, x^1 = r, x^2 = \theta, x^3 = \phi$. Now, one can compute the $R_{\mu \nu}$ for the orthonomal basis given in eq.(8) as follows:
\begin{equation}
R_{00} = e^{-2\nu} ( \mu '' + {\mu'}^{2} - \mu ' \nu ' + \frac{2 \mu'}{r} )
\end{equation}
\begin{equation}
R_{11} = e^{-2 \nu} ( -\mu '' + {\lambda'}^2 + \mu ' \nu ' + \frac{2 \nu '}{r} )
\end{equation}
\begin{equation}
R_{22} = R_{33} =\frac{1}{r^2} ( 1 - e^{-2 \nu}) + \frac{1}{r} ( -\mu'+ \nu') e^{- 2 \nu} \end{equation}
For  static electrically charged solutions, the only non-zero field strength is  $F_{r t}$ or $F_{01}$. This leads to,
\begin{equation}
\frac{\partial L(F)}{\partial g^{11}} = -\frac{\partial L(F)}{\partial g^{00}}; \hspace{1.0cm} \frac{\partial L(F)}{\partial g^{22}} = \frac{\partial L(F)}{\partial g^{33}} = 0. 
\end{equation}
Note that the conditions in equation(12) are satisfied by any function $L$ of $F^2$ for static electrically charged metric.
The condition in eq.(12) and the fact that  $\eta_{ab} = -,+,+,+$, combined with the above field equations  leads to,
\begin{equation}
R_{11} = -R_{00}\hspace{1.0cm} \Rightarrow \mu + \nu = 0
\end{equation}
In order to solve for the function $\mu$ and $\nu$, we will use the  electromagnetic equations(4). The non-vanishing components of the electrodynamic field tensor in the coordinate basis are given by $F_{tr} = E(r)$ and in the orthonormal basis $F_{01} = \hat{E}$. They are related to each other by,
$E(r) = \hat{E}(r) e^{\mu +\nu}$.
Hence, the non-linear Lagrangian reduces to,
\begin{equation}
 L(F) = 4 \beta^2 \left( 1 - \sqrt{ 1 - \frac{E^2}{\beta^2}} \right)
\end{equation} 
Note that the eq.(14) imposes an upper bound for $|E|$ to be smaller than $\beta$. This is a crucial characteristic of non-linear electrodynamics which leads to finite self energy of the electron. Th electric field $E$ follows from eq.(4) as,
\begin{equation}
E(r)  = \frac{Q}{\sqrt{ r^4 + Q^2/\beta^2}}
\end{equation}
From Einstein's equations one can derive a relation for $\nu$ as follows,
\begin{equation}
( r e^{-2 \nu})' = 1 - \Lambda r^2 + 2\beta ( r^2 \beta - \sqrt{ Q^2 + r^4 \beta^2} )
\end{equation}
Hence,
\begin{equation}
e^{-2 \nu} = 1 - \frac{2M}{r} - \frac{ \Lambda r^2}{3} + 2 \beta \left( \frac{r^2 \beta}{3} - \frac{1}{r} \int_r ^{\infty} \sqrt{ Q^2 + r^4 \beta^2} \right)
\end{equation}
In the limit $\beta \rightarrow \infty$, the elliptic integral can be expanded to give,
\begin{equation}
(re^{-2 \nu})' = 1 - \Lambda r^2 - \frac{ Q^2}{r^2}
\end{equation}
resulting  the function $\nu(r)$ for the Reissner-Nordstrom-(anti)de Sitter solutions,
\begin{equation}
e^{- 2 \nu} = 1 - \frac{2M}{r} - \frac{\Lambda r^2}{3} + \frac{ Q^2}{r^2}
\end{equation}
Here $M$ is an integrating constant which may be interpreted as a quasi-local mass when $\Lambda =0$ \cite{zan}.

\section{Electrically  Charged Black Hole Solutions}

In this section we will explore the existence of black holes solutions 
from the above general solutions. For regular horizon to exist the 
metric function $g_{tt}= e^{-2 \nu}$ has to have zeros. First,
let us redefine $e^{-2 \nu}$ as $f(r)$ and define $g_0(r)$ as the function $d(rf(r))/dr$. 
$$ f(r) = 1 - \frac{2M}{r} - \frac{ \Lambda r^2}{3} + 2 \beta \left( \frac{r^2 \beta}{3} - \frac{1}{r} \int_r^{\infty} \sqrt{ Q^2 + r^4 \beta^2}\right)
$$
\begin{equation}
g_0(r) = 1 + ( 2 \beta^2 - \Lambda ) r^2 - 2 \beta \sqrt{ Q^2 + \beta^2 r^4}
\end{equation}
The number of positive roots of $f(r)$ and of $rf(r)$ are the same. Therefore, the number of roots of $f(r)$ would be at most  the number of roots of $g_0(r)$ plus one. (which is a result of Rolle's theorem in calculus). Since we are looking for roots of $g_0(r)$, we will consider the function $g(r)$ which is obtained by simplifying $g_{0}(r) = 0$.
\begin{equation}
g(r) =  r^4 ( \Lambda^2 - 4 \beta^2  \Lambda ) + r^2 ( 4 \beta^2  - 2 \Lambda ) + (1 - 4 \beta^2 Q^2)
\end{equation}

To understand the behavior of the black hole metric for very small $r$, one can do a series expansion of $f(r)$ around $r = 0$ as follows:
\begin{equation}
f(r) \approx 1 - \frac{( 2M - A)}{r} - \frac{10}{3} \beta Q + \frac{2 \beta^2}{3} r^2  - \frac{ \beta^2}{5}  r^4
\end{equation}
Here,
\begin{equation}
A = \frac{1}{3} \sqrt{ \frac{\beta}{ \pi} } Q^{3/2} \Gamma \left(\frac{1}{4} \right)^2
\end{equation}
We will define $2M' = 2M - A$ to discuss the behavior of the geometry.
Hence when $r \rightarrow 0$, the behavior of $f(r)$ is dominated by $2M'/r$ term irrespective of the value of $Q$ and $\Lambda$. One may recall that in the Einstein-Maxwell black hole solutions, it is $Q^2/r^2$  that dominates the behavior of the space-time around $r=0$. This is a major difference incorporating Born-Infeld electrodynamics to gravity compared to the Maxwell case.

We will classify the solutions according to the value of the integration constant $M'$ in the following sections. We will consider $\Lambda > 0$ and $\Lambda <0$ cases separately. Note that the $\Lambda = 0$ case has been studied by Rasheed \cite{ras}.

\subsection{ De Sitter Black Holes}
For $\Lambda >0$,  $g(r)$ is a 4th degree even function and can have at the most two positive roots. Hence $f(r)$ can have at the most three roots. However, if one takes a closer look at the behavior of the function for $r \rightarrow \infty$ and $r \rightarrow 0$ one can predict the number of roots exactly. From  the expansions in eq.(22), it is obvious that for small $r$, $f(r) \rightarrow  -2M'/r$ and for large $r$,  $f(r) \rightarrow - \infty$. 

When  $M' > 0$, for  small $r$, the function $f(r) \rightarrow - \infty$. Hence $f(r)$ does not have a root at all or it has two roots or it has a degenerate root as shown in the Fig.1 . In all these cases, the curvature scalars $ R_{\alpha \gamma\beta \rho} R^{\alpha \gamma \beta \rho}$, $R_{\alpha \beta} R^{\alpha \beta}$ and $R$ explode at $r = 0$ and are finite else where. Hence there is a curvature singularity at $r =0$. Since for large $r$, $f(r)< 0$, $r$ becomes a time coordinate. Hence the future infinity is space-like and such space-times have cosmological event horizons. Such space-times are  similar to  Schwarzschild-de Sitter black hole space-time \cite{hawkin}.

In the Fig.1, the graph in plain line represents  a solution with two roots to the function $f(r)$. The smaller of these values, which can be called $r_{+}$ can be regarded as the position of the black hole event horizon. The larger value of the roots $r_{++}$ represents the position of the cosmological event horizon for observers on the world-lines of constant $r$ between $r_+$ and $r_{++}$. The Killing vector $\frac{\partial}{\partial t}$ is null on both these horizons. These horizons can be removed by appropriate coordinate patches similar to the Schwarzschild-de-Sitter black hole space-time \cite{hawkin}.

As $M'$ increases, the black hole event horizon increases and the cosmological event horizon decreases. For special values of the $M'$, the two coincides leading to a degenerate horizon as given in the Fig.1 (dashed).

\begin{center}
\scalebox{.9}{\includegraphics{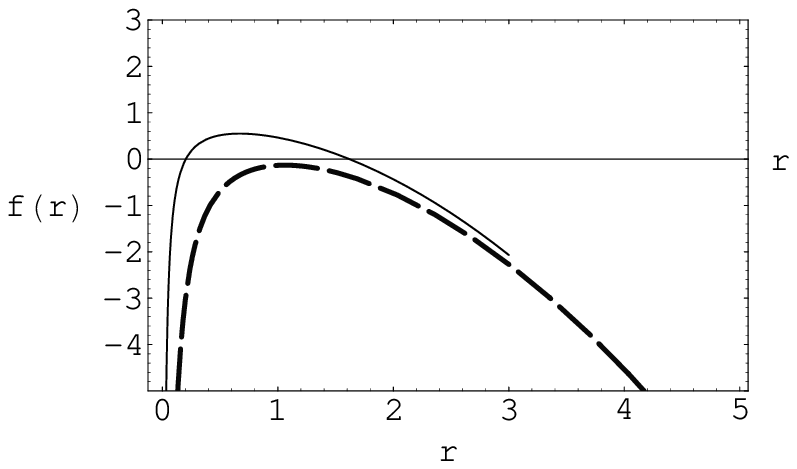}}

\vspace{0.3cm}
\end{center}
Figure 1. The figure shows the function $f(r)$  for $\Lambda=1.5$, $\beta=0.01$  and $Q=1$. The lighter graph shows $f(r)$ for $M'=0.1$ and the dashed one is for $M'=0.4$.

\subsection{ Anti de-Sitter Black Holes}
For $\Lambda < 0$, one can apply Decarte's rule on the discriminant to predict the behavior of the roots of $g(r)$ as follows; Since $ (2 \beta^2 - \Lambda) > 0$, $g(r)$ can have only one root leading $f(r)$ to have two roots.

When $M'  > 0$  the function $f(r) \rightarrow  -\infty$ for small $r$ and goes to 
$ \infty$ for large $r$.
From the Fig. 2, there is only one  horizon. There is a curvature singularity at $ r =0$. Hence this is very similar to Schwarzschild black hole with a space-like singularity at the origin.

\begin{center}
\scalebox{.9}{\includegraphics{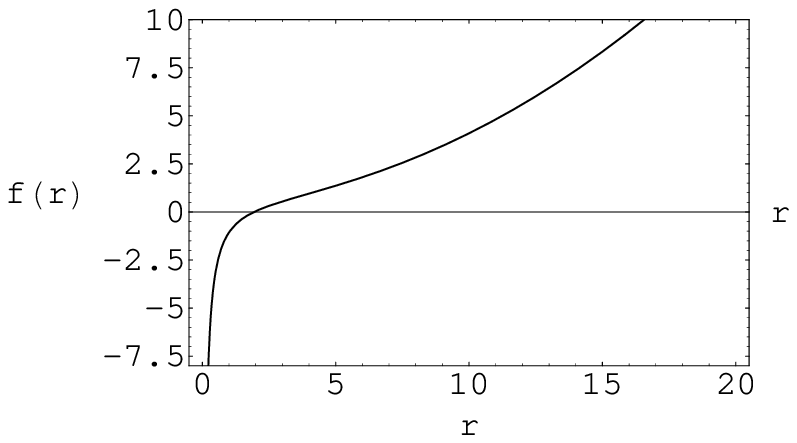}}

\vspace{0.3cm}
\end{center}
Figure 2. The figure shows the function $f(r)$  for $\Lambda=-0.1$, $\beta=0.05$, $M'=1$  and $Q=1$

When  $M' = 0$ the function $f(r) \rightarrow  (1 - 10\beta Q/3)$ fo small $r$ and
goes to $ \infty$ for large $r$.
From the Fig. 3, the function $f(r)$ can have one root or none. However, there is no curvature singularity at $ r =0$ and the function $f(r)$ is bounded at $r=0$. Hence this is  a particle-like solution.

\begin{center}
\scalebox{.9}{\includegraphics{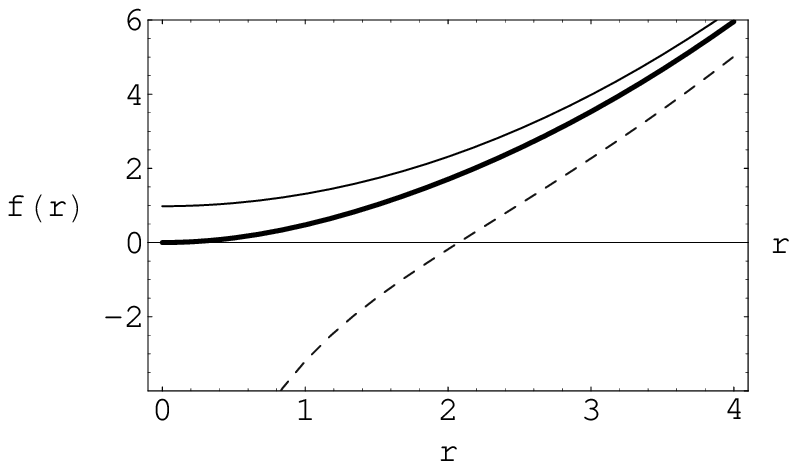}}

\vspace{0.3cm}
\end{center}
Figure 3. The figure shows the function $f(r)$  for $\Lambda=- 1$, $M'=0$  and $Q=1$. The plain graph is for $\beta = 0.01$, the heavy is for $\beta = 0.5$ and the dashing is for $\beta = 5$

\section{Extreme Black Holes}

Here we will also discuss the extreme black hole in detail since it is a counterpart to the BPS state in Reissner-Nordstrom solutions.
For degenerate roots of $f(r)$, both $f(r)$ and $df(r)/dr$ are zero. Hence from eq.(16), one arrives at the following equation for horizon radii $r_{ex}$:
\begin{equation}
r_{ex}^4 (\Lambda^2 - 4 \beta^2 \Lambda) + r_{ex}^2 (-2 \Lambda + 4 \beta^2 ) + (1 - 4Q^2 \beta^2) = 0
\end{equation}
which gives the solution as,
\begin{equation}
(r^2_{ex})_{ \pm} = \frac{ ( 2\beta^2 - \Lambda) \pm \sqrt{\delta }}{ \Lambda ( 4 \beta^2 - \Lambda)}
\end{equation}
Here,
\begin{equation}
\delta = (2\beta^2 - \Lambda)^2 + ( 1 - 4\beta^2 Q^2) ( 4\beta^2 - \Lambda) \Lambda 
\end{equation}
The existence of  $r_{ex}$  varies for various values of the parameters of the theory as follows:

For $\Lambda < 0$, $(4 \beta^2 - \Lambda) > 0$,  the denominator of $r_{ex}$ is negative.  Since $(2\beta^2 - \Lambda) >0$, there is at most one positive root for $r_{ex}^2$. It occurs only if $(1 -4 \beta^2 Q^2)( 4 \beta^2 - \Lambda) \Lambda >0$ leading to   $\beta Q < 1/2$. The corresponding value is given by $r^2_{ex+}$ of eq.(24).

For $ 2\beta^2 > \Lambda >0$, the denominator of eq.(24) is positive and so is the first element of the numerator. Thus there is always one positive root. A second will occur if $(1 - 4\beta^2 Q^2) < 0$ while $\delta$ remains positive.

For $ 4 \beta^2 > \Lambda > 2 \beta^2 $,  the denominator of eq.(24) is positive, while the first element of the numerator is negative. There is a positive root  if 
$\delta >0$ i.e. $\beta Q < 1/2$.

For $ \Lambda > 4 \beta^2$, both the denominator and the first element of the numerator are negative. So there is always a positive root and there will be two if $(\delta - (2 \beta^2 - \Lambda)^2) < 0$  ( or $\beta Q >1/2$) and $\delta$ remains positive.

\section{Temperature}
The Hawking temeperature of the black hole solutions discussed above can be calculated as follows: $T = {\kappa}/2 \pi$. Here $\kappa$ is the surface gravity given by 
$$ \kappa = - \frac{1}{2} \frac{dg_{tt}}{dr} |_{r= r_{+}} $$
Here, $r_+$ is the event horizon of the black hole. Since $f(r) = 0$ at $r = r_+$, the eq.(16) can be used to calculate the surface gravity exactly and the corresponding temeparature as,
\begin{equation}
T = \frac{1}{4\pi} \left( \frac{1}{r_+} - \Lambda r_+ + 2\beta \left( r_+ \beta - \frac{\sqrt{(Q^2 + r^2 \beta^2)}}{r_+} \right) \right)
\end{equation}

\section{Conclusions}

In this paper we presented  the local properties of the solutions arising in Einstein-Born-Infeld gravity with a cosmological constant. 
The non-singular nature of the electric field at the origin changes the structure of  the space-times drastically. The singularity at the origin is dominated by the mass term $M/r$ rather than the $Q^2/r^2$. Hence the number of horizons changes when compared to the Einstein-Maxwell gravity.
Currently we are studying the global properties of these space-times in detail.

In extending this work, we would like to study the stability of these black holes in detail. The electrically charged black holes in anti-de Sitter space have been shown to be unstable for large black holes by using linear perturbation techniques \cite{gub}. It would be interesting to study how the non-linear nature effects the instability of such solutions.

From a string theoretical point of view, it is vital to study  black hole solutions arising in Einstein-Born-Infeld-dilaton gravity. It is well known that the presence of a dilaton changes the space-time drastically. In such as case, the Lagrangian to be considered would be a more general one with a $SL(2,R)$ invariant Born-Infeld term with a dilaton-axion coupled to it. Solitons and black hole solutions were constructed in a recent paper by Clement and Gal'tsov for Einstein-Born-Infeld-dilaton gravity without  a cosmological constant \cite{clem}.
It would be interesting to study the effects of both the Born-Infeld term and the dilaton with a cosmological constant which we hope to report elsewhere.

One may recall that the Reissner-Nordstrom AdS black hole is supersymmetric for $Q=0$ and for $Q^2 = M^2$ which was shown by Romans \cite{rom}. It would be interesting to embed the black hole solutions obtained in this paper in a supergravity theory.\\

{\bf Note:} The references  \cite{will} and \cite{car} was added after the paper was published. 
\\

{\bf Acknowledgement} This work was supported by the grant number: 2001-16 of the Center for Integrative Natural Science and Mathematics of Northern Kentucky University.

\end{document}